\documentstyle[12pt]{article}
\begin{document}
\tolerance=5000
\def\pp{{\, \mid \hskip -1.5mm =}}
\def\cL{{\cal L}}
\def\be{\begin{equation}}
\def\ee{\end{equation}}
\def\bea{\begin{eqnarray}}
\def\eea{\end{eqnarray}}
\def\tr{{\rm tr}\, }
\def\nn{\nonumber \\}
\def\e{{\rm e}}
\def\D{{D \hskip -3mm /\,}}

\def\SEH{S_{\rm EH}}
\def\SGH{S_{\rm GH}}
\def\AdS5{{{\rm AdS}_5}}
\def\S4{{{\rm S}_4}}
\def\gfv{{g_{(5)}}}
\def\gfr{{g_{(4)}}}
\def\SC{{S_{\rm C}}}
\def\RH{{R_{\rm H}}}

\def\wlBox{\mbox{
\raisebox{0.1cm}{$\widetilde{\mbox{\raisebox{-0.1cm}\fbox{\ }}}$}}}
\def\htBox{\mbox{
\raisebox{0.1cm}{$\hat{\mbox{\raisebox{-0.1cm}{$\Box$}}}$}}}


\  \hfill 
\begin{minipage}{3.5cm}
August 2001 \\
\end{minipage}

\vfill

\begin{center}
{\large\bf Determinant Line Bundles and Topological
Invariants of Hyperbolic Geometry - Expository Remarks}

\vfill

{\sc Andrei A. BYTSENKO}\footnote{email: abyts@uel.br},
{\sc Marcos C. FALLEIROS}\footnote{email: faleiros@npd.uel.br},
{\sc Antonio E. GON\c CALVES}\footnote{email: goncalve@uel.br}, 
and {\sc Zhanna G. KUZNETSOVA}\footnote{email: jane@fisica.uel.br}, 
\\

\vfill

{\sl Departamento de Fisica, Universidade Estadual de Londrina,\\ 
Caixa Postal 6001, Londrina-Parana, Brazil}

\vfill


{\bf ABSTRACT}

\end{center}

{\small 
We give some remarks on twisted determinant line bundles and Chern--Simons
topological invariants associated with real hyperbolic manifolds.
Index of a twisted Dirac operator is derived. We discuss briefly application
of obtained results in topological quantum field theory.}


\newpage

\section{Introduction}

The topological invariants have been explicitly calculated for a 
number of 3-manifolds and gauge groups
\cite{dijk90-129-393,kirb91-105-473,free91-141-79,jeff92-147-563,rama93-8-2285,
roza93u-99,roza96-175-275}. 
In dimension three there are two important topological quantum field theories
of cohomological type, namely topological $SU(2)$ gauge theory of flat 
connection and a version of the Seiberg-Witten theory. The twisted 
${\mathcal N}=4$ $SUSY\,\,\, SU(2)$ pure gauge theory (version of the 
Donaldson-Witten
theory) describes the Casson invariant \cite{blau93} while Seiberg-Witten 
theory is a 3d twisted version of ${\mathcal N}=4$ $SUSY\,\,\, U(1)$ gauge 
theory with matter multiplet \cite{seib94,witt94}. The both theories can be 
derived from 4d ${\mathcal N}=2$ $SUSY\,\,\,SU(2)$ gauge theory 
\cite{Odintsov} corresponding 
via twist to Donaldson-Witten theory. It would be interesting and natural to 
investigate dual description of the ${\mathcal N}=2$ theory in low-energy
limit. It could provides formulation of invariants of four-manifolds involving
elements of the Chern-Simons invariants. 

The Chern-Simons invariant ${W}_{CS}(X;k)$ weightted by 
$\exp(\sqrt{-1}kCS(A))$ has to all orders in
$k^{-1}=\hbar/2\pi$ \,\,$(k\in {\bf Z})$ an asymptotic stationary phase 
approximation of the form \cite{roza96}:

$$
{W}_{CS}(X;k)=\sum_{j}{W}^{(j)}(X;k)
\exp\,\sqrt{-1}k\left(CS(A^{(j)})+\sum_{n=2}^{\infty}CS_n(A^{(j)})k^{-n}
\right)
\mbox{,}
\eqno{(1.1)}
$$
\\
where $CS(A^{(j)})$ is the Chern-Simons secondary characteristic class 
related to the flat connections $A^{(j)}$ and $CS_n(A^{(j)})$ are the 
n-loop quantum corrections coming from the n-loop 1-particle irreducible 
Feynman diagrams. 

The semiclassical approximation for the 
Chern-Simons partition function (quadratic functional) may be expressed by
the asymptotics for  $k\rightarrow\infty$ of Witten's invariant. This 
asymptotics leads to a series of
$C^{\infty}-$ invariants associated with triplets $\{X;F;\xi\}$ with $X$ a 
smooth homology 
$3-$ sphere, $F$ a homology class of framings of $X$, and $\xi$ an acyclic 
conjugacy class of ortogonal representations of the fundamental group
$\pi_1(X)$ 
\cite{axel94-39-173}. In addition the cohomology $H(X;Ad\,\xi)$ of $X$ 
with respect to the local system related to $Ad\,\xi$ vanishes. 

Our aim here are designed to provide readers with a brief introduction
to the anomaly and index formulas, associated with Chern--Simons
invariants and to indicate how these formulas
are derived. We shall limit our discussion, in particular,
to the real hyperbolic manifolds. 

We begin with some conventions which apply throughout. 
Let $X$ be a locally symmetric Riemannian manifold with negative sectional 
curvature. Its universal covering ${\widetilde   X}\rightarrow X$ is a 
Riemannian symmetric space of rank one.
The group of orientation preserving isometries ${\widetilde G}$ of 
${\widetilde   X}$ is a
connected semisimple Lie group of real rank one and 
${\widetilde   X}={\widetilde   G}/{\widetilde   K}$,
where ${\widetilde   K}$ is a maximal compact subgroup of ${\widetilde   G}$. 
The fundamental group of $X$ acts by covering transformations on 
${\widetilde   X}$ and gives
rise to a discrete, co-compact subgroup $\Gamma \subset {\widetilde   G}$ 
such that $X=\Gamma\backslash {\widetilde   G}/{\widetilde   K}$. 
If $G$ is a linear connected finite 
covering of ${\widetilde   G}$, the embedding 
$\Gamma\hookrightarrow {\widetilde   G}$ lifts
to an embedding $\Gamma\hookrightarrow G$. Let $K\subset G$ be a maximal
compact subgroup of $G$, then $\Gamma\backslash G/K$ is a compact
manifold. For $G=SO(m,1)$ \,$(m\in {\bf Z}_{+})$, $K=SO(m)$. 
The corresponding symmetric space of non-compact type is the real hyperbolic
space ${\bf H}^m$ of sectional curvature $-1$. It's compact dual space is
the unit $m-$ sphere.
In Sect. 3 we shall be working with real compact hyperbolic 
manifolds with topology 
$\Gamma\backslash G/K = \Gamma\backslash {\bf H}^{2n}$ (for details see Refs. 
\cite{eliz,byts96-266}).

\section{Determinat lines and eta invariants}

The differential geometry of determinant line bundles 
has been developed in \cite{quillen} in a special case and in 
\cite{bismut1,bismut2} in general. 
In papers \cite{dai,freed} the results on 
eta invariants (the exsponentiated eta invariant naturally takes values in 
the determinant line of the boundary) 
were used to reprove the holonomy formula for determinant line 
bundles, known as Witten's global anomaly formula \cite{witten}.

Results which we have considered in the 
paper hold for any Dirac operator on a ${\rm Spin}^{\bf C}$ manifold
coupled to a vector bundle with connection (manifolds are assumed spin
and its metric is a product near the boundary).
On a closed odd dimensional manifold $X$ the Dirac operator ${\cal D}_X$ is 
self-adjoint and has discrete spectrum ${\rm Spec}\,({\cal D}_X)$. The 
eta invariant 
$\eta_X(s)$ of Atiyah-Patodi-Singer \cite{atiyah} by general estimates is 
converges for $\Re\,(s)$ sufficiently large. For Dirac operators the 
meromorphic
continuation is analytic for $\Re\,(s)>-2$ \cite{bismut2}. At $s=0$ the eta 
invariant is reguliar and we set

$$
\tau_X=\exp\left(\pi\sqrt{-1}(\eta_X(0,{\cal D}_X)+{\rm dim}\,{\rm Ker}\,
{\cal D}_X)
\right)\in{\bf C}
\mbox{.}
\eqno{(2.1)}
$$
\\  
The eta invariant $\eta_X(0,{\cal D}_X)$ is discontinuous in general 
but the general theory shows 
that $\tau_X$ varies smoothly in families ($|\tau_X|=1$).

We recall some standard material on Dirac bundles
(for details see, for example, Ref. \cite{laws89}). Let $\partial X$ be 
an even-dimensional orientable Riemannian manifold, and let 
${\bf E}\rightarrow {\partial X}$ be a hermitian vector bundle equipped
with a compatible connection $\nabla$. Let ${C}\ell(\partial X)$ denote 
the complexified Clifford bundle. We suppose that there is a bundle map from
${C}\ell(\partial X)\rightarrow {\rm End}\,{\bf E}$ which is an algebra 
homomorphism on each fiber and which covers the identity:
\vspace{1cm}

$$
\begin{picture}(120,80)
\put(60,30){$\partial X$}
\put(0,80){${C}\ell(\partial X)$}
\put(100,80){${\rm End}\,{\bf E}$}
\put(40,83){\vector(1,0){52}}
\put(28,75){\vector(1,-1){30}}
\put(105,75){\vector(-1,-1){30}}
\put(210,50){$$}
\end{picture}
\eqno{(2.2)}
$$
\\
is commutative. Denote by ${\cal S}$ the spin bundles associated to the two
half-spin representation of ${\rm Spin}\,({\rm dim}\,\partial X)$; then 
${\cal S}={\cal S}^{+}\oplus{\cal S}^{-}$ where ${\cal S}^{\pm}$ are 
half-spin bundles. $\nabla$ induces a connection on 
${\cal S}\otimes{\bf E}$ which is compatible with both the metric and the 
${\bf Z}_2-$ grading. The latter connection canonically defines a Dirac 
operator ${\cal D}_{\partial X}: C^{\infty}({\cal S}^{+}\otimes{\bf E})
\rightarrow C^{\infty}({\cal S}^{-}\otimes{\bf E})$ described by

$$
{\cal D}_{\partial X}: C^{\infty}({\cal S}_{\bf E}^{+})\stackrel
{\nabla}{\longrightarrow}
C^{\infty}(T^{*}\partial X\otimes {\cal S}_{\bf E}^{+})\stackrel{{\cal H}}
{\longrightarrow}C^{\infty}({\cal S}_{\bf E}^{-})
\mbox{,}
\eqno{(2.3)}
$$
\vspace{0.3cm}

\noindent 
where ${\cal H}: T^{*}\partial X\longrightarrow 
{\rm Hom}({\cal S}^{+}_{\bf E},
{\cal S}^{-}_{\bf E})$ denotes the Clifford multiplication and 
${\cal S}^{\pm}_{\bf E}\stackrel{def}{=}{\cal S}^{\pm}\otimes{\bf E}$.
Then ${\rm dim}\,{\rm Ker}^+{\cal D}_{\partial X}={\rm dim}\,
{\rm Ker}^-{\cal D}_{\partial X}$ and the boundary condition is an isometry
$I: {\rm Ker}^+{\cal D}_{\partial X}\rightarrow {\rm Ker}^-{\cal D}_
{\partial X}$. Therefore the invariant (2.1) is defined and we have 
$\tau_X \in {\rm Det}_{\partial X}^{-1}$,
where ${\rm Det}_{\partial X}$ is the determinant line of the Dirac operator 
${\cal D}_{\partial X}$ on the boundary: ${\rm Det}_{\partial X}=
({\rm Det}\,{\rm Ker}^-{\cal D}_{\partial X})
\otimes ({\rm Det}\,{\rm Ker}^+{\cal D}_{\partial X})^{-1}$.

Let a family of Riemannian manifolds is a smooth fiber bundle 
$\pi: X\rightarrow Z$ together with a metric on the relative tangent 
bundle $T(X/Z)$
which endowed with spin structure. Also we asume that the Riemannian metrics
on the fibers are products near boundary. 
The determinant lines carries the Quillen metric and a canonical connection
$\nabla$ \cite{bismut1} and the exponentiated eta invariant is a smooth 
section $\tau_{X/Z}: Z\rightarrow {\rm Det}_{{\partial X}/Z}^{-1}$.

We mention here two basic results on this invariant, namely variation and
curvature formulas.
\medskip
\medskip
\par \noindent
{\bf Theorem 2.1.}\,\,\,(Dai and Freed \cite{dai}, Theorem 1.9.)\,
{\em The covariant derivative of the exponentiated eta invariant is

$$
\nabla \tau_{X/Z} = 2\pi\sqrt{-1}\left[\int_{X/Z}\widehat{A}
(\Omega^{X/Z})\right]_{(1)}\cdot \tau_{X/Z}
\mbox{,}
\eqno{(2.4)}
$$
where $\Omega^{X/Z}$ is the Riemannian curvature of $X\rightarrow Z$,
$\widehat{A}$ is the usual polynomial

$$
\widehat{A}(\Omega) = \sqrt{{\rm det}\left(\frac{\Omega/4\pi}{{\rm sinh}\,
\Omega/4\pi}
\right)}
\mbox{,}
\eqno{(2.5)}
$$
${\rm Tr}e^{\sqrt{-1}\Omega/2\pi}={\rm ch}(\Omega)$ and simbol (1) 
denotes the 1-form 
piece of the differential form.} 
\\
\\
\\
Note that for a family of closed manifolds this is a result of 
Atiyah-Patodi-Singer.
The fiber of $X\rightarrow Z$ we denote as $X/Z$ and let it will be 
a closed manifold. Let $KO(X)$ is the group of virtual real vector 
bundles on $X$ 
up to equivalence and 
$KO^{-N}(X)\subset KO({\bf S}^N \otimes X)$, \, $N={\rm dim}(X/Z)$, is 
the subspace of isomorphism classes of virtual bundles trivial on ${\bf S}^N 
\bigvee X$. The spin structure on the fibers can be determined by a 
pushforward map $\pi_{!}^{X/Z}: KO(X)\rightarrow KO^{-N}(Z)$.
Let $E\rightarrow X$ is a real vector bundle; the family of Dirac operators on
$X/Z$ coupled to $E$ has an index in $KO^{-N}(Z)$, which equals 
\cite{atiyah76} $\pi_{!}^{X/Z}([E])$, where $[E]\in KO(X)$ is the isomorphism
class of $E$.

In general for Dirac operators coupled to complex bundles in K-theory one can 
express the Chern character of the index in terms of Chern character of 
$E$ by means of the formula \cite{freed99}:

$$
{\rm ch}\,\pi_{!}^{X/Z}([E])=\pi_{*}^{X/Z}({\hat A}(X/Z){\rm ch}\,(E))
\mbox{,}
\eqno{(2.6)} 
$$
\\         
where $\pi_{*}$ is the pushforward map in rational cohomology.

Let the fibers $X/Z$ are closed. Then the determinant line bundle 
${\rm Det}\,{\cal D}^{X/Z}(E)$ is a well-defined as a smooth line bundle
(it carries a canonical metric and connection) \cite{bismut86}. The complex
Dirac operator for the odd dimensional fibers $X/Z$ is self-adjoint and 
there is a geometrical invariant $\xi_{X/Z}(E): Z\rightarrow 
{\bf R}/{\bf Z}$ defined by Atiyah-Patodi-Singer,

$$
2\xi_{X/Z}=(\eta(0)+{\rm dim}{\rm Ker}\,{\cal D})_{X/Z}
\mbox{.}
\eqno{(2.7)}
$$
\\
We have

$$
\tau_{X/Z}(E)=\exp\left(2\pi\sqrt{-1}\xi_{X/Z}\right): 
\,\,\, Z\rightarrow \bf T
\mbox{,}
\eqno{(2.8)}
$$
\\
where $\bf T \subset \bf C$. 
\medskip
\medskip
\par \noindent
{\bf Theorem 2.2.}\,\,\,(Bismut and Freed \cite{bism86-107}, Theorem 1.21.)\,
{\em The 2-form curvature of the determinant line bundle
is 

$$
\Omega^{{\rm Det}\,{\cal D}^{X/Z}(E)}=\left[2\pi\sqrt{-1}\int_{X/Z}{\hat A}
(\Omega^{X/Z}){\rm ch}\,(\Omega^{E})\right]_{(2)}\in\Omega^2(Z)
\mbox{,}
\eqno{(2.9)}
$$
where $\Omega^{X/Z}$ and $\Omega^{E}$ are the curvature forms.}
\\
\\

Let $X\rightarrow {\bf S}^1$ is a loop of manifolds in 
this geometric setup. A metric and spin structure on $X$ could be induced by
a metric and boundary spin structure on ${\bf S}^1$. The holonomy of the
determinant line bundles around loop takes the form

$$
{\rm hol}\,{\rm Det}\,{\cal D}^{X/{\bf S}^1}(E)={\rm a}-{\rm lim}\,
\tau_{X}^{-1}(E)
\mbox{,}
\eqno{(2.10)}
$$
\\
where a-lim is the adiabatic limit, i.e. the limit as the metric on 
${\bf S}^1$ blows up ($\varepsilon\rightarrow 0:\,\, {\rm g}_{\bf S}^1
\rightarrow {\rm g}_{\bf S}^1\varepsilon^{-2}$). For the flat determinant 
line 
bundles no adiabatic limit is required. As we pointed out before, 
Eq. (2.10) is the 
global anomaly formula \cite{witten,bismut86}.

\section{Index of the Dirac operator}

Calculating index of the Dirac operator we shall follow the lines of paper 
\cite{barbasch83} and consider special 
case $G=SO_1(2n,1)$, \,$K=SO(2n)$. The complexified Lie algebra 
${\rm g}={\rm g}^{\bf C}_0={so}(2n+1,{\bf C})$ 
of $G$ is of Cartan type $B_n$ with Dynkin diagram

$$
\underbrace{\bigcirc-\bigcirc-\bigcirc \cdots \bigcirc-\bigcirc}_{2n\,
{\rm  nodes}} 
= \bigcirc
\mbox{.}
\eqno{(3.1)}
$$
\\
The standard systems of positive roots 
$\triangle^{+},\triangle^{+}_k$ for ${\rm g}$ and $k=k^ {\bf C}_0$ -
the complexified Lie algebra of $K$, with respect to a Cartan subgroup
$H$ of $G$,\, $H\subset K$, are given by 

$$
\triangle^{+}=\{\varepsilon_i|1\leq i\leq n\}\bigcup \triangle^{+}_k
\mbox{,}
\eqno{(3.2)}
$$
\\
where 

$$
\triangle^{+}_k=\{\varepsilon_i\pm \varepsilon_j|1\leq i<j\leq n\}
\mbox{,}
\eqno{(3.3)}
$$
\\
and

$$
\triangle^{+}_n \stackrel{def}{=}\{\varepsilon_i|1\leq i\leq n\}
\mbox{,}
\eqno{(3.4)}
$$
\\
is the set of positive non-compact roots. Here

$$ 
(\varepsilon_i, \varepsilon_j)=\frac{\delta_{ij}}{2(2n-1)}
\mbox{,}
\eqno{(3.5)}
$$
\\
$(\,,\,)$ being the Killing form of G. Let ${h}_0$ be the 
Lie algebra of $H$ and let 
${h}^{*}_{\bf R}={\rm Hom}(\sqrt{-1}{h}_0,{\bf R})$ be the dual 
space of the real vector space
$\sqrt{-1}{h}_0$. Thus the $\{\varepsilon_i\}_{i=1}$ are an 
${\bf R}-$ basis 
of ${h}^{*}_{\bf R}$. Of interest are the {\em integral} elements 
$f$ of ${h}^{*}_{\bf R}$:

$$
f\stackrel{def}{=}\{\mu\in {h}^{*}_{\bf R}|\frac{2(\mu,\alpha)}
{(\alpha,\alpha)}\in 
{\bf Z},\,\,\, \forall \alpha\in \triangle^{+}\}
\mbox{.}
\eqno{(3.6)}
$$
\\
Using Eq. (3.5) we have

$$
\frac{2(\mu,\varepsilon_i)}{(\varepsilon_i,\varepsilon_i)}
=2\mu_i\,\,\,\,\,\,\,\,\,\,\,\,\,\,\,\,\,\,\,\,\,\,\,\,\,\,\,\, 
{\rm for}\,\,\,\,\, 1\leq i\leq n
\mbox{,}
\eqno{(3.7)}
$$
\\
$$
\frac{2(\mu,\varepsilon_i\pm \varepsilon_j)}
{(\varepsilon_i\pm \varepsilon_j, \varepsilon_i\pm \varepsilon_j)}
=\mu_i\pm \mu_j \,\,\,\,\, {\rm for}\,\,\,\,\,1\leq i<j\leq n
\mbox{,}
\eqno{(3.8)}
$$
\\
where we shall write $\mu=\sum_{j=1}^n\mu_j\varepsilon_j$ for $\mu\in 
{h}^{*}_{\bf R},\, \mu_j\in {\bf R}$. Then clearly
\\
$$
f=\left\{\mu\in {h}^{*}_R|\,\,\,\,\,\,\,
2\mu_i\,\,\in \,{\bf Z}\,\,\,\,\,\,\,\,\,\,\,\,\,\,\,\,\,\,
{\rm for}\,\,\,1\leq i\leq n \right.
$$
$$
\left.
\:\:\:\:\:\:\:\:\:\:\:\:\:\:\:\:\:\:\:\:\:\:\:\,\,
\:\:\:\:\:\:\:\:\:\:\:\:\:\:\:\:\:\:\:\:\:
\mu_i\pm \mu_j \in \,{\bf Z}\,\,\,\,\,\,\,\,\,\,\,\,\,\,\,\,\,
{\rm for}\,\,\,\,1\leq i<j\leq n \right\}
\mbox{.}
\eqno{(3.9)}
$$
\\

Let $\delta_k=\frac{1}{2}\sum_{\alpha\in\triangle^{+}_k}\alpha, \,\,
\delta_n=\frac{1}{2}\sum_{\alpha\in\triangle^{+}_n}\alpha, \,\, 
\delta=\delta_k+\delta_n=\frac{1}{2}\sum_{\alpha\in\triangle^{+}}\alpha$.
Then $\delta_k=\sum_{i=1}^n(n-i)\varepsilon_i,\,\,
\delta_n=\frac{1}{2}\sum_{i=1}^n\varepsilon_i,\,\,
\delta=\sum_{i=1}^n(n-i-\frac{1}{2})\varepsilon_i$ are all integral.
Elements $\mu$ of $f$ corresponds to characters $e^{\mu}$ of $H$.
Following the paper \cite{barbasch83} we fix once and for all $\mu\in f$ 
which is $\triangle^{+}_k-$ dominant: $(\mu,\alpha)\geq 0$ for 
$\alpha\in\triangle^{+}_k$. For us, in concrete terms, this means the 
following: let $\mu=(\mu_1,...,\mu_n)$ be a sequence of real numbers
such that

\vspace{0.6cm}
({\bf i}) $2\mu_i\in {\bf Z}$ for $1\leq i\leq n$ and $\mu_i\pm \mu_j \in 
{\bf Z}$ for $1\leq i<j<n$ (i.e. $\mu$ 

is integral).

({\bf ii}) $|\mu_n|\leq\mu_{n-1}\leq\mu_{n-2}\leq...\leq\mu_2\leq\mu_1$
and either every $\mu_i\in {\bf Z}$ or 

every $\mu_i$ is half an odd integer.

\vspace{0.6cm}
Note that in fact $({\bf ii})\rightarrow ({\bf i})$ so that we may drop 
condition $({\bf i})$ (only for $G=SO_1(2n,1)$ since in general $({\bf ii})$
does not imply $({\bf i})$). Then $\mu+\delta_n$ does define a character of 
$H$ as has been required in \cite{barbasch83} and by their construction 
($\mu$ be the
highest weight of an irreducible $k-$ module $V_{\mu}$) there is a twisted 
Dirac operator ${\cal D}^{\chi}_{\Gamma\backslash X}$ on a vector bundle over
$\Gamma\backslash G/K$ for $\Gamma$ a discrete subgroup of $G$ for $\Gamma$
satisfying the conditions of their paper.

Thus $\Gamma\backslash G$ need not be compact, but one requires of course that
${\rm Vol}(\Gamma\backslash G)<\infty$. 
The paper \cite{barbasch83} also
requires that a choice $\psi$ of positive root system for 
$({\rm g},{h}={h}^{\bf C}_0)$ be chosen such that
$\triangle^{+}_k\subset \psi$
and such that $\mu+\delta_k$ is $\psi-$ dominant, i.e.
$(\mu+\delta_k,\alpha)\geq 0\,\,\forall\,\,\alpha\in \psi$ (condition
(1.3.1) of \cite{barbasch83} where 
$\rho_c=\delta_k,\,\,\Phi=\triangle\stackrel{def}{=}
\triangle^{+}\bigcup(-\triangle^{+}),\,\,\Phi_k=\triangle^{+}_k,\,\,\Phi_n=
\triangle^{+}_n,\,T=H$). Here we choose 
$\psi=\triangle^{+}$ provided $\mu_n\geq 0$, which we now assume; see
$({\bf ii})$.

\subsection{Twisted Dirac operator}
In summary we assume the following. $\mu=(\mu_1,...,\mu_n)$ is a sequence of
real numbers, and 

\vspace{0.6cm}
$({\bf iii})\,\,0\leq\mu_n\leq\mu_{n-1}\leq\mu_{n-2}
\leq\cdot\leq\mu_2\leq\mu_1$, where either every 

$\mu_j\in {\bf Z}$ or every 
$\mu_j$ has the form $\mu_j=n_j+\frac{1}{2}$ for some $n_j\in {\bf Z}$.
\\
\\
\
Then we have a twisted Dirac operator 
${\cal D}^{\chi}_{\Gamma\backslash X}$ over
$\Gamma\backslash G/K$.

\medskip
\medskip
\par \noindent
{\bf Theorem 3.1} (Barbasch and Moscovici \cite{barbasch83} for the case 
$G=SO_1(2n,1), n\geq 2$). \,\,\,
{\em For a suitable normalization of Haar measure on $G$,
and for $\mu$ satisfying condition ({\bf iii}) one has}

$$
{\rm Index}\,{\cal D}^{+}_{\mu,\Gamma}={\rm Vol}(\Gamma\backslash G)
\frac{\prod_{\alpha\in \triangle^{+}}(\mu+\delta_k,\alpha)}
{\prod_{\alpha\in \triangle^{+}_k}(\delta_k,\alpha)}
\mbox{.}
\eqno{(3.10)}
$$
\\
\\
To make this explicit we must express $P_k\stackrel{def}{=}
\prod_{\alpha\in\triangle^{+}_k}(\delta_k,\alpha)$ and
$P\stackrel{def}{=}\prod_{\alpha\in\triangle^{+}}(\mu+\delta,\alpha)$ directly
in terms of the real numbers $\mu_1,...,\mu_n$. 

Tedious calculation gives
the final result for $\mu=(\mu_1,...,\mu_n)\in {\bf R}^n$ subject to
condition $({\bf iii})$ (for suitable Haar measure on $G$). Namely,
the $L^2-$ index of twisted Dirac operator ${\cal D}^{+}_{\mu,\Gamma}$ 
is equal to
\\
\\
$$
{\rm Index}\,{\cal D}^{+}_{\mu,\Gamma}={\rm Vol}(\Gamma\backslash G)
\frac{P}{P_k}=
\frac{{\rm Vol}(\Gamma\backslash G)}{[2(2n-1)^n]}
$$
$$
\times
\frac{\prod_{i=1}^n(\mu_i+n-i)\prod_{1\leq i<j\leq n}(\mu_i+\mu_j+2n-i-j)
(\mu_i-\mu_j-i+j)}{\prod_{1\leq i<j\leq n}(2n-i-j)(-i+j)}
\mbox{.}
\eqno{(3.11)}
$$
\\

\section{Concluding remarks}

There are many applications of these formulae. We note some of
them. For example, the results obtained in the paper could be used 
for calculation of Chern--Simons functional related to a real hyperbolic
3-manifold  
\cite{bytsenko97,bytsenko98,bytsenko99,bytsenko99l,bytsenko99p,bytsenko00}:
\\
$$
CS(A_{\chi})=\frac{1}{2\pi\sqrt{-1}}{\rm log}\left[
\frac{Z(0,{\cal D})^{{\rm dim}\,\chi}}{Z(0,{\cal D}_{\chi})}\right]
+({\rm dim}\,\chi\,{\rm Index}\,{\cal D}_{{A}_{{\chi}_0}}
-{\rm Index}\,{\cal D}_{{A}_{\chi}})
$$
\\
$$
\!\!\!\!\!\!\!\!\!\!\!\!\!\!\!\!\!\!\!\!\!\!\!\!
\!\!\!\!\!\!\!\!\!\!\!\!\!\!\!\!
+\frac{1}{2}\left({\rm dim}\,\chi\,
{\rm dim}{\rm Ker}\,{\cal D} -
{\rm dim}{\rm Ker}\,{\cal D}_{\chi}\right)_{\Gamma\backslash {\bf H}^3}
\mbox{.}
\eqno{(4.1)}
$$
\\
Here $Z(s,{\cal D}_{\chi})$ is a Selberg type (Shintani) zeta
function \cite{millson,mosc} and any 1-dimensional representation 
$\chi$ of $\Gamma$
factors through a representation $H^1(X,{\bf Z})$, and $\chi_0$ is a
trivial representation. 

Eq. (2.10) is 
known as the global anomaly formula \cite{witten,bismut86}. 
Global anomalies in the worldsheet path integral of string theory in the
presence of $D-$ branes has ben studied in \cite{freed99}. Recently agreement 
between partition functions of IIA string theory and $M-$ theory 
(a derivation $K-$ theory from $M-$ theory) was found in \cite{diac}. 
One can calculate also the phase of the one-loop
contributions from instantons to the membrane superpotential \cite{harvey}. 
In all of these edxamples the lolonomy of the determinant line bundles 
(and the index of a Dirac operator, for example (3.11)) play very 
important role.

\section*{Acknowledgements}

A.A.B. very grateful to S.D. Odintsov and F.L. Williams for helpful 
discussion.
The work of A.A.B. was supported in part by the Russian Foundation for
Basic Research (grant No. 01-02-17157). A.E.G. thanks CNPq for partial 
support.

\end{document}